# Microstructure Formation in Freezing Nanosuspension Droplets

*Mathieu Nespoulous\*, Renaud Denoyel , and Mickaël Antoni*

Aix-Marseille Université, CNRS, MADIREL UMR 7246, Marseille, France

**Corresponding Author**

mathieu.nespoulous@univ-amu.fr

**ABSTRACT**

The structural evolution of suspensions upon freezing is studied with optical microscopy in a suspended droplet configuration. Droplets have millimeter size and consist in an aqueous mixture of silica particles while the surroundings phase is hexane. Freeze-thaw cycles are applied to this system and a two-step freezing mechanism evidenced. A fast adiabatic growth of dendrites that invade the full droplets is first observed, and occurs within a few milliseconds. Then a slow process lasts for several seconds and corresponds to the release of solidification latent heat into the hexane phase. The striking feature of this work is to evidence that after the first freeze-thaw cycle flocculated microstructures are generated. When a second cycle is performed, microstructures further flocculate and generate, for dense silica suspensions, stable porous spheres of the size of the droplets. A phenomenological description based on repulsion or engulfment of particles by solidifying ice fronts is proposed.

**TOC GRAPHICS**

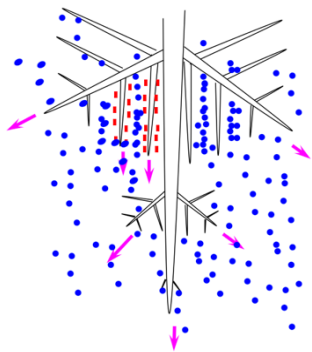

The solidification dynamics of undercooled liquids is of fundamental interest for many areas of science such as materials science [1], metallurgy [2], cloud physics [3, 4] or food preservation [5]. In the case of colloidal suspensions, it also has critical importance in engineering processes such as the so-called freeze-casting [6, 7] or freeze drying [8, 9]. In this process, the solid/liquid interface propagates through the colloidal suspension during solidification and its interaction with the particles can lead, after sublimation of the frozen phase, to the formation of porous materials [10, 11, 12]. Considering the freezing of pure water, one particular aspect is the link between dendritic growth and latent heat release. It is known that dendrites can grow with velocities up to meters per seconds [13]. This is far too fast to allow solidification latent heat to be evacuated. For undercooled droplets, two successive freezing regimes are therefore naturally expected and observed [14, 15]. Firstly, dendrites are formed. They can invade the whole medium but with a volume fraction limited by the undercooling temperature. The heat provided by the solidification is indeed reheating the droplets up to the melting point of water in a quasi-adiabatic process. Secondly, driven by heat exchange with the cold surrounding medium, an ice front advances from the droplet interface to its center at a speed imposed by the heat conduction properties. As shown recently [15], the kinetics of this second regime depends upon the surface to volume ratio of the droplets and, for very small ones, the speed of the two regimes may become comparable.

In the present work, the objective is to investigate this problem when water is replaced by a colloidal suspension of silica particles. The following questions will be addressed: (i) are the aforementioned two regimes still observable, (ii) are the characteristic dynamic time scales modified and (iii) what is the incidence of freezing processes on the suspension stability and on its structural evolution?

To answer these fundamental questions, a system consisting in droplets pending at the end of capillary tip is used (see figure 1). Droplets are made of an aqueous silica nanoparticles suspension and immerged in hexane. When submitting this system to freeze-thaw cycles, an unexpected aggregation process where silica microstructures or even a full caking of the droplets can be evidenced. Scanning electron microscopy (SEM) experiments, performed on the freeze-dried nanofluid droplets after two cycles, reveal the formation of a porous spherical particle with size similar to that of the original droplet and exhibiting macropores. An explanation for this phenomenon based on the balance of the electrostatic interactions of the nanoparticles and ice surface charge is proposed. It is to be noted that a similar phenomenology has been evidenced when replacing hexane by air. Hexane is mainly used for the optical purpose of refractive index matching with water.

When cooling pure water droplets below water melting temperature, the freezing occurs in two subsequent regimes at an undercooling temperature around -18°C in the present experiments. During the first regime, a dendritic growth phenomenon takes place as illustrated in figure 1a. It is fast since within a few milliseconds the droplet is completely filled with a three-dimensional network of entangled dendrites. As nucleation starts in most cases at the capillary tip, this network is growing top-bottom. This first regime will be denoted in the following as dendritic-freezing regime (DFR). The measured dendritic front velocity is about 0.15 m/s, in agreement with recent results [15]. The second regime, corresponds to the bulk-freezing regime (BFR) where a front of ice moving from the droplet interface to its center (see figure 1b) completely solidifies the droplet. It results from the quasi-equilibrium of isothermal bulk-water solidification and is mainly driven by the heat propagation in the entangled medium built up by the boundaries of the dendrites. In BFR, freezing front velocity is about $10^{-4}$ m/s.

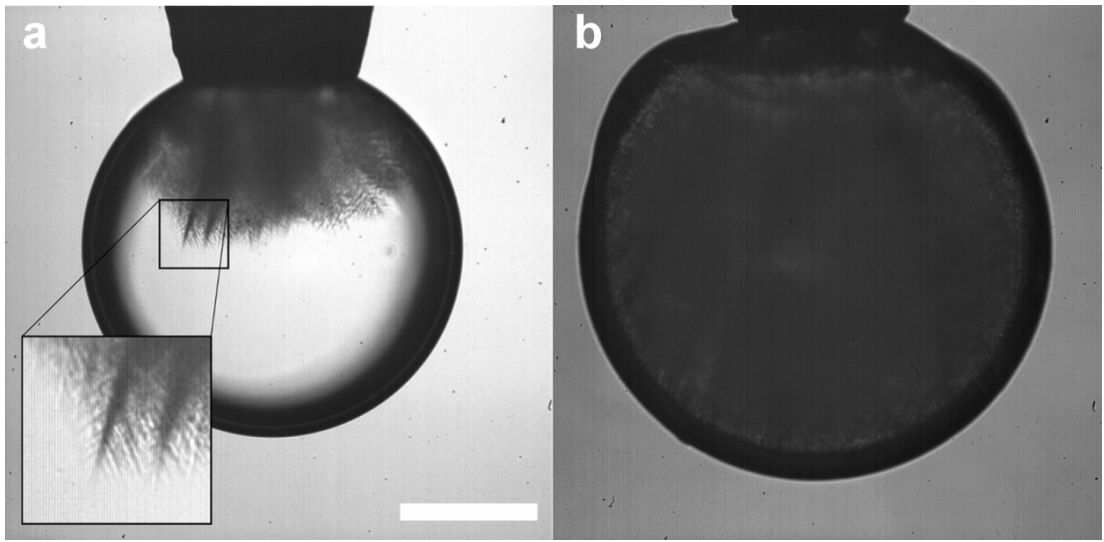

**Figure 1.** a) Optical microscopy image of DFR of a pure water droplet (cooling rate 1°C/min). Scale bar is 250 µm. Inset: zoom in the droplet showing the dendrite network. b) Shape of the solidification front in BFR. In the black layer at the droplet/hexane interface water is frozen while liquid water is still present elsewhere.

Various concentrations of silica nanoparticles in water have been analyzed (from 3 to 300 g/L). A concentration of 300 g/L in silica nanoparticle is equivalent to 11.3 vol.%. Initially, all the droplets are transparent indicating the absence of light diffusing objects. Snapshots of the DFR are given in figure 2 for two silica concentrations. Transparent area at the bottom of the droplets indicate that DFR is still running. BFR (not shown here) shows the same features as for pure water. Figure 2 clearly indicates that the topology of dendrites depends upon the particle concentration, with a more regular dendrite front when silica concentration increases. The velocities $v_{DFR}$ and $v_{BFR}$ of DFR and BFR fronts are displayed in figure 3 as a function of silica nanoparticles concentration. $v_{DFR}$ takes values close to 0.15 m/s for most concentrations but drops down to 0.08 m/s for [$SiO_2$] = 300 g/L. This can be explained by the increase of the suspension viscosity when concentrated [16]. The data for $v_{BFR}$ are scattered but bounded within $5.0\ 10^{-5}$ m/s and $1.5\ 10^{-4}$ m/s. After the first freeze-thaw cycle, droplets recover their initial shape but surprisingly turn turbid. This is a typical signature of the presence of microstructures. When kept at room temperature for several hours after the first melting, a layer of sedimented solid material is observable at the bottom of the droplets. DLS analyses of the supernatant phase indicate the presence of particles with an average diameter of 120 nm and a dispersity parameter value of 0.25.

The formation of microstructures from the original nanoparticles suspension demonstrates that freezing triggers an irreversible aggregation process of the nanoparticles. When applying a second freeze-thaw cycle right after the first one, the same phenomenology shows up (first a DFR followed by a BFR). Unexpectedly, the final product, if silica concentration is 300 g/L, is a porous silica pearl with a diameter close to that of the original droplet, completely textured by water freezing. The mechanical strength of this pearl is actually low but high enough to resist the capillary pressure during drying at room temperature.

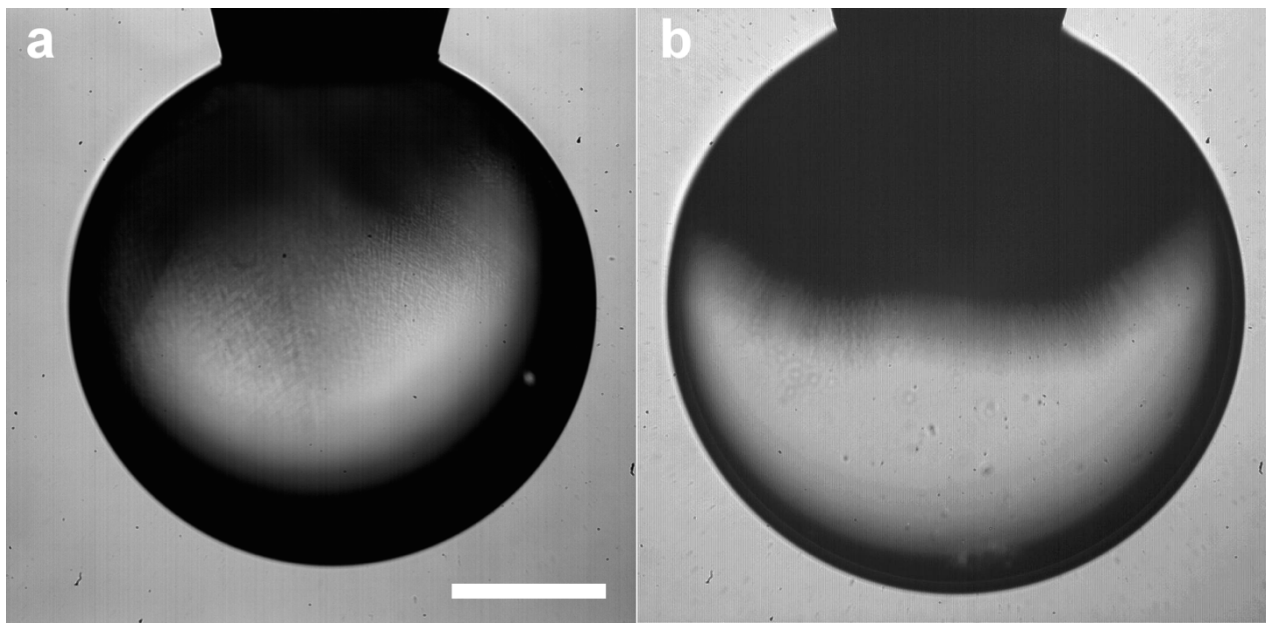

**Figure 2.** Optical micrograph of suspended nanofluid droplets in DFR (cooling rate 1°C/min) for [SiO2] = 3 g/L (a) and [SiO2] = 300 g/L (b). Scale bar is 250 µm.

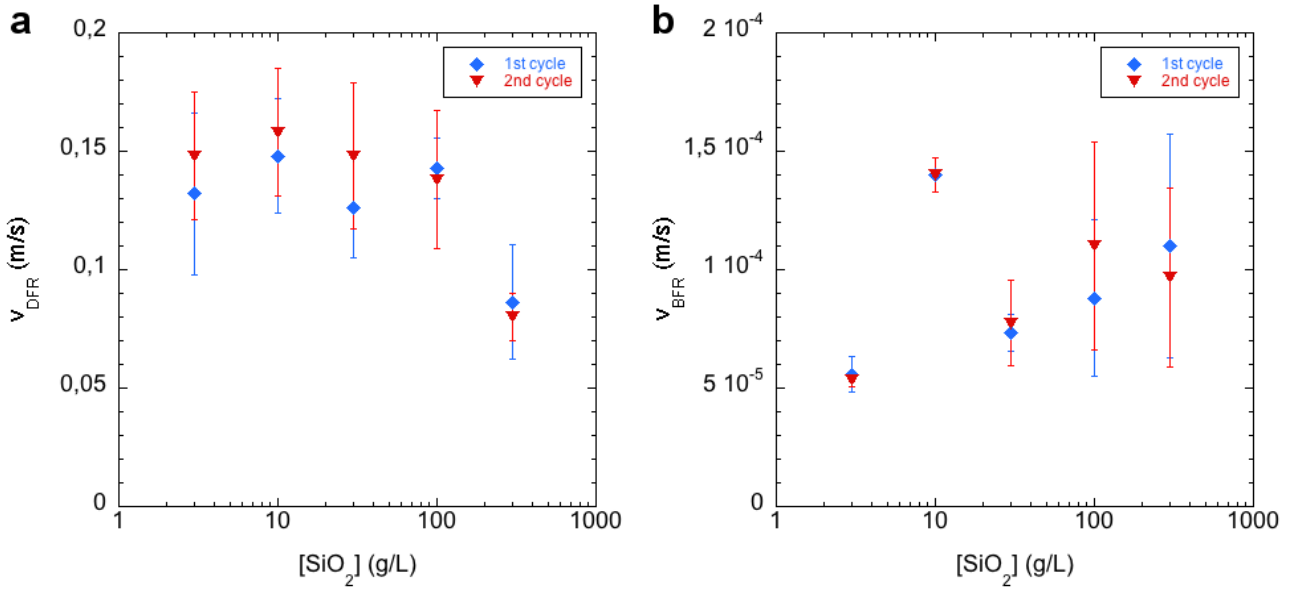

**Figure 3.** (a) (resp. (b)) $v_{DFR}$ (resp. $v_{BFR}$) as a function of [SiO2]. Diamonds (resp. triangles) correspond to the velocities of the first (resp. second) freeze-thaw cycle.

Scanning electron microscopy analysis of the porous pearls of Figure 4 reveal a complex structure where nanoparticles define a three-dimensional macropore network. The morphology of the porous network seems to be different from what is obtained by freeze casting with a low unidirectional freezing rate [17, 18], which mainly leads to one dimensional structures. Here, the topology of the textures shows up as connected cells that resemble to the ones evidenced in cryo-SEM analysis of freeze-fractured nanofluid droplets [19]. The typical size of the visible pores in this micrograph is about 5 µm. They can be seen as signature of dendrite growth although their organization is not as regular as in directional freezing [10]. Beside this macroporosity, the walls are most probably mesoporous with a pore size of 8 nm given by the one of a compact arrangement of spherical nanoparticles. This was estimated from nitrogen adsorption at 77 K and application of the BJH method to a sample obtained by drying a large volume of the suspension.

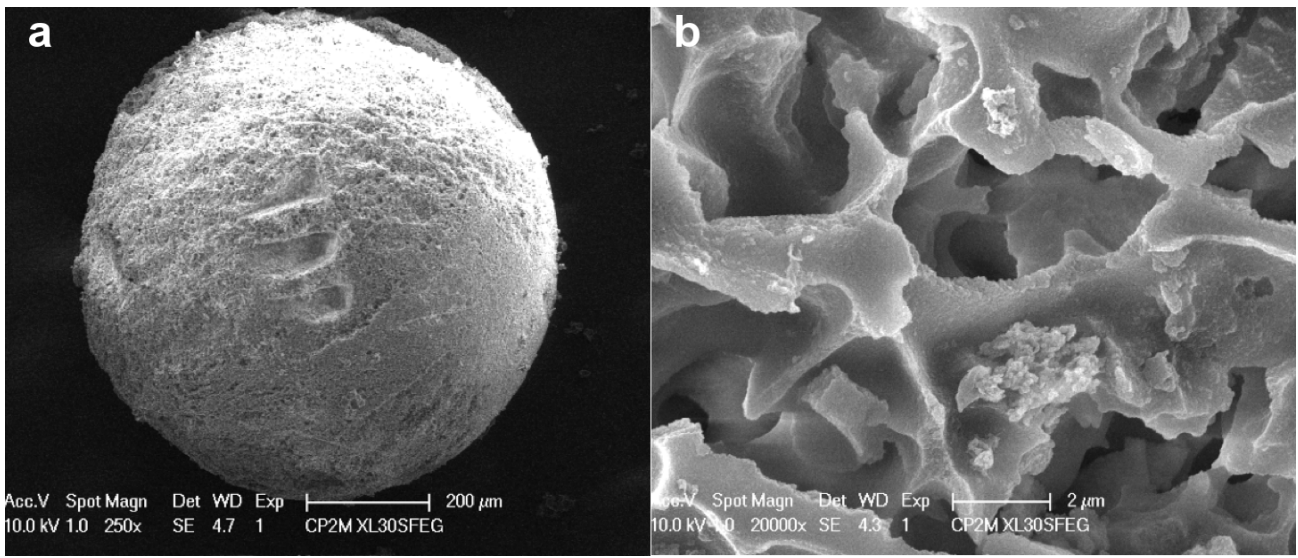

**Figure 4.** a) Scanning electron microscopy pictures of a freeze-dried droplet ([SiO2] = 300 g/L). Scale bar is 200 µm. b) Zoom-in of (a) showing the porous structure of the medium. Scale bar is 2 µm.

To understand the observed behavior, the interactions between the ice front and the nanoparticles have to be accounted for. Both van der Waals and electrostatic contributions must therefore be considered here [20]. Four main interactions have to be accounted for in DFR and BFR: (i) The disjoining force between ice and nanoparticles which may be attractive or repulsive depending on the Hamaker constant. (ii) The electrostatic force between charged nanoparticles and charged ice front. This is a repulsive force in the present case since both silica and ice surface are negatively charged [21]. (iii) The DLVO forces that operate between nanoparticles and create an osmotic pressure which acts against the increase of nanoparticle concentration when the ice invades the suspension. (iv) Finally, the viscous forces that hinder the displacement of nanoparticles [16]. Buoyancy forces are negligible in this problem, moreover, freeze-thaw cycles are performed within short time scales (about 15 minutes), giving no time to sedimentation to set in. The surface potential of ice (point (ii)) can actually be split into two contributions. The first one is the reversible potential due to the ionization of OH groups at the surface of ice. It is considered to be around -25 mV in the experimental conditions of this work [21, 22]. The second is the freezing potential that develops at the liquid/solid interface for freezing water. It appears when the propagation of liquid/solid fronts is fast and results from the fact that anions and cations are not engulfed at the same rate [23]. This potential is known to decrease both with ionic strength and ice front speed [24, 25].

Following the analysis of Azouni et al [16], a critical velocity, $v_C$, can be introduced for the ice front: below this velocity, particles are repelled, above they are engulfed. $v_C$ is defined as the velocity at which the sum of the forces acting between particles and ice front is zero. Most of these forces are repulsive except the viscous force which is opposite to particle displacement. The equations of the model are detailed in the Supporting Information; and the values of the parameters corresponding to the silica suspension under focus here are shown in Table 1. The behavior of $v_C$ with respect to particle radius is displayed in Figure 5.

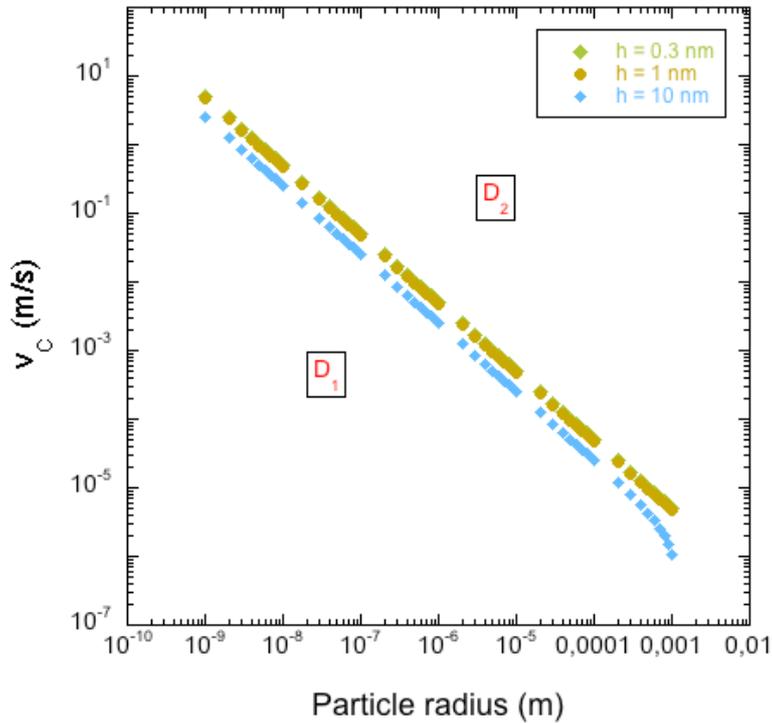

**Figure 5.** $v_C$ as a function of particle radius. In domain $D_1$, $v<v_C$, and particles are pushed by the ice front while in $D_2$, $v>v_C$ and particles are engulfed in ice. The legend indicates the distance h between ice front and particles.

One fundamental parameter in the model for the computation of $v_C$ is the distance h between the ice front and particles. Figure 5 displays the value of $v_C$ for h = 0.3 nm, 1.0 nm and 10 nm. Despite this broad range of h values, $v_C$ shows only a weak dependency with this parameter when r < 1 mm. This allows the definition of two distinct domains $D_1$ and $D_2$ (below and above the curve respectively). In $D_1$ particles are pushed along the propagating front while in $D_2$ they are engulfed [17].

Initially, the nanoparticles in the suspension have typical radius r = 15 nm. The critical velocity is then $v_C$ = 0.5 m/s and Figure 3 indicates that $v_C>v_{DFR}>v_{BFR}$ for the first freeze-thaw cycle. Both DFR and BFR are hence in $D_1$ and nanoparticles therefore repelled by the ice front. After DFR, the suspension concentration in the unfrozen part is increased by 20% [14]. As a result, in the subsequent BFR, nanoparticles experience growing repulsive forces as they operate in more and more confined conditions. Dendritic network is indeed already fully developed and droplet volume most probably partitioned in ice free interstitial domains. As BFR goes on, these latter are reached by the freezing front and the suspension within them concentrates more and more until capillary pressure becomes comparable to osmotic pressure [5]. Conditions are then reached to overcome the repulsive electrostatic barrier and flocculation to be triggered. The result here is the production of aggregates in the micrometric range which leads to a suspension with completely different granulometry.

After this first freeze-thaw cycle, the typical size of the particles, that are now microparticles, has been increased by two orders of magnitude. According to Figure 5, this means that $v_C$ drops down to about $10^{-2}$ m/s. Figure 3 demonstrates that $v_{DFR}$ and $v_{BFR}$ are almost unchanged when performing a second cycle. This means that the mechanisms involved in particle flocculation can be studied along a horizontal line (resp. horizontal strip) located at 0.15 m/s for the new DFR (resp. $(1.0 \pm 0.5)\ 10^{-4}$ for the new BFR) in figure 5. DFR situation is now drastically changed. As the suspension is micrometer sized, DFR occurs in $D_2$ and particles are, depending on their size, engulfed or partially engulfed by dendrites but not pushed as in the first cycle. In BFR however, $v_{BFR} < v_C$ showing that engulfment is still not favored. Hence, similarly the first freeze-thaw cycle, microparticles are compressed against each other in a pre-existing complex frozen dendrite network. However, conversely to what happened in this first cycle, the compression between microstructures now occurs in a 3-dimensional dendrite network that itself contains a relevant fraction of engulfed aggregates. These latter, as BFR takes place, are brought into contact with the microstructures already trapped in the dendrite free interstitial domains and create a new generation of flocculated objects. For large $SiO_2$ concentrations, this scenario yields a massing of the suspension that finally turns into a single porous pearl (see Figure 4).

In the same configuration of single, isolated droplet, methods are now being developed to produce large amounts of such pearls with the goal to carry out experimental campaigns of mercury porosimetry and gas adsorption. This well controlled protocol could also be seen here as new and original route for the synthesis of porous materials.

This article focuses on the freezing dynamics of silica nanoparticle suspensions in a suspended droplet configuration. A two-step scenario is evidenced where the solidification first exhibits a fast dendritic growth followed by bulk freezing. Solidification velocity is measured and shown to depend upon the nanoparticle concentration. When performing freeze-thaw cycles the structure of the dispersion is strongly modified since from nanodispersed it turns microdispersed. To explain this scale change, the interplay between surface charge of solidification fronts in the liquid/solid coexisting system and interparticle repulsive forces is addressed. Flocculation mechanisms generated by particle confinement in a dendrite network are shown to occur. After two freeze-thaw cycles of the droplets, initially dense suspensions are massing and form a single flocculated pearl. This work demonstrates that granulometry of suspensions strongly evolve under freezing and raises the fundamental question of the consequence of this phenomenon in the microphysics of atmospheric clouds when air contains fine particles.

**Table 1**: Parameters and constants used in the model.

| Description | Parameter (unit) | Value [ref.] |
|---|---|---|
| Radius of np* | R (m) | $1.5 \times 10^{-8}$ |
| np - ice front distance | $h_0$ (m) | $1 \times 10^{-9}$ |
| Boltzman constant | $k_B$ (J.K$^{-1}$) | $1.38 \times 10^{-23}$ |
| Gravity constant | g (m.s$^{-2}$) | 9.81 |
| Working temperature | T (K) | 273.15 |
| Curvature parameter | $\alpha$ | 0 [16] |
| Density of water @ 10°C | $\rho_L$ (kg.m$^{-3}$) | 1000 |
| Density of water @ 0°C | $\rho_S$ (kg.m$^{-3}$) | 917 |
| Density of np | $\rho_P$ (kg.m$^{-3}$) | 2648 |
| Viscosity of water @ 0°C | $\mu$ (Pa.s) | $1.79 \times 10^{-3}$ [26] |
| Mass concentration of np | c (kg.m$^{-3}$) | 0.3 |
| Debye length | $\lambda_D$ (m) | $9.1 \times 10^{-9}$ |
| Inverse Debye length | $\kappa = 1/\lambda_D$ (m$^{-1}$) | $1.1 \times 10^{8}$ |
| Surface potential | $\Psi_0$ (V) | 0.03 |
| np-water-np Hamaker const. | $A_{p\text{-}p}$ (J) | $7.46 \times 10^{-21}$ [27] |
| np-water(ice) Hamaker const. | $A_{p\text{-}i}$ (J) | $-8.2 \times 10^{-22}$ [16] |

| Electric constant of vacuum | $\varepsilon_0$ ($C^2.J^{-1}$) | 8.85x10$^{-12}$ [26] |
| --- | --- | --- |
| Dielectric constant of water @ 0°C | $\varepsilon_W$ ($C^2.J^{-1}$) | 87.9 [26] |
| Surface charge of np-water | $\sigma_S$ ($C.m^{-2}$) | -0.02 [28] |
| Surface charge of ice-liq. interface | $\sigma_P$ ($C.m^{-2}$) | -0.01 [21] |

*np stands for silica nanoparticle.

Experimental Methods

**Sample preparation** - Aqueous solutions of amorphous silicon oxide ($SiO_2$) nanoparticles (Levasil 200, AkzoNobel) were used. The $SiO_2$ nanoparticles have a diameter of 30 nm. They are negatively charged and show a good stability when dispersed into water (with a zeta potential of about -40 mV). Stock solutions of 30% mass fraction in colloids are used.

A microliter droplet of the colloidal dispersion is formed within a vial previously filled with hexane. The alkane acts as a continuous surrounding phase for the pending droplet.

**Freezing experiment** - The pending droplet solidification dynamics is analyzed with optical microscopy. A Peltier cooling system allows to cycle temperature at a given cooling rates (usually 1°C/mn). The microscope is equipped with a CMOS camera (Mikrotron MC1310). The typical freeze-thaw cycle applied on the droplet is: from room temperature to -20 °C to room temperature, at a constant 1 °C/mn rate.

**SEM experiment** – A previously described procedure was followed to analyze the pearls

[19]. The scanning electron microscope used is a Philips XL30 SFEG STEM associated to an Oxford EDS analysis system.

**Supporting Information.** Detailed expression of the forces used in the model to compute the critical velocity.

# REFERENCES

[1] Zhang, H. F.; Hussein, I.; Brust, M.; Butler, M. F.; Rannard, S.P.; Cooper, A. I. Aligned Two- and Three-dimensional Structures by Directional Freezing of Polymers and Nanoparticles. *Nature Mater.* **2005**, *4*, 787–793.

[2] Li, J. F.; Liu, Y. C.; Lu, Y. L.; Yang, G. C.; Zhou, Y. H. Structural Evolution of Undercooled Ni-Cu Alloys. *J. Cryst. Growth*. **1998**, *192*, 462-470.

[3] Bogdan, A.; Molina, M. J.; Tenhu, H.; Mayer, E.; Loerting, T. Formation of Mixed-phase Particles During the Freezing of Polar Stratospheric Ice Clouds. *Nature Chem*. **2010**, *2*, 197-201.

[4] Gultepe, I.; Tardif, R.; Michaelides, S. C.; Cermak, J.; Bott, A.; Bendix, J.; Müller, M. D.; Pagowski, M.; Hansen, B.; Ellrod, G.; et al. Fog Research: A Review of Past Achievements and Future Perspectives. *Pure Appl. Geophys*. **2007**, *164*, 1121-1159.

[5] Olivera, D. F.; Salvadori, V. O. Effect of Freezing Rate in Textural and Rheological Characteristics of Frozen Cooked Organic Pasta. *J. Food Eng*. **2009**, *90*, 271-276.

[6] Deville, S.; Maire, E.; Bernard-Granger, G.; Lassale, A.; Gauthier, C.; Leloup, J.;

Guizard, C. Metastable and Unstable Cellular Solidification of Colloidal Suspensions. *Nature Mater*. **2009**, *8*, 966-972.

[7] Li, W. K.; Lu, K.; Walz, J. Y. Freeze Casting of Porous Materials: Review of Critical Factors in Microstructures Evolution. *Int. Mater. Rev*. **2012**, *57*, 37-60.

[8] Han, J.; Zhou, C.; Wu, Y.; Liu, F.; Wu, Q. Self-assembling Behavior of Cellulose Nanoparticles During Freeze-drying: Effect of Suspension Concentration, Particle Size, Crystal Structure, and Surface Charge. *Biomacromolecules*. **2013**, *14*, 1529-1540.

[9] Vicent, M.; Sanchez, E.; Molina, T.; Nieto, M. I.; Moreno, R. Comparison of Freeze Drying and Spray Drying to Obtain Porous Nanostructured Granules from Nanosized Suspensions. *J. Eur. Ceram. Soc*. **2012**, *32*, 1019-1028.

[10] Deville, S.;Saiz, E.; Nalla, R. K.; Tomsia, A. P. Freezing as a Path to Build Complex Composites. *Science*. **2006**, *311*, 515-518.

[11] Shanti, N. O.; Araki, K.; Halloran, J. W. Particle Redistribution during Dendritic Solidification of Particle Suspensions. *J. Am. Ceram. Soc*. **2006**, *89*, 2444-2447.

[12] Thongprachan, N.; Nakagawa, K.; Sano, N.; Charinpanitkul, T.; Thantapanichakoon, W. Preparation of Macroporous Solid Foam from Multi-walled Carbon Nanotubes by Freeze-drying Technique. *Mat. Chem. Phys*. **2008**, *112*, 262-269.

[13] Hobbs, P. Ice Physics; Oxford University Press: London, 1974.

[14] Hindmarsh, J. P.; Russel, A. B.; Chen, X. D. Experimental and Numerical Analysis of the Temperature Transition of a Suspended Freezing Water Droplet. *Int. J. Heat Mass Transfer*. **2003**, *46*, 1199-1213.

[15] Buttersack, T.; Bauerecker, S. Critical Radius of Supercooled Water Droplets: on the Transition toward Dendritic Freezing. *J. Phys. Chem. B*. **2016**, *120*, 504-512.

[16] Azouni, M. A.; Casses, P. Thermophysical Properties Effects on Segregation during Solidification. *Adv. Colloid Interface Sci*. **1998**, *75*, 83-106.

[17] Chino, Y.; Dunand, D. C. Directionally Freeze-cast Titanium Foam with Aligned, Elongated Pores. *Acta Mater*. **2008**, *56*, 105-113.

[18] Peppin, S. S. L.; Wettlaufer, J. S.; Worster, M. G. Experimental Verification of Morphological Instability in Freezing Aqueous Colloidal Suspensions. *Phys. Rev. Lett*. **2008**, *100*, 238301.

[19] Limage, S.; Schmitt, M.; Vincent-Bonnieu, S.; Dominici, C.; Antoni, M. Characterization of Solid-stabilized Water/Oil Emulsions by Scanning Electron Microscopy. *Colloids Surf. A*. **2010**, *365*, 154-16.

[20] Stahlberg, J.; Appelgren, U.; Jonsson, B. Electrostatic Interactions between a Charged Sphere and a Charged Planar Surface in an Electrolyte Solution. *J. Colloid Interface Sci*. **1995**, *176*, 397-407.

[21] Kallay, N.; Cakara, D. Reversible Charging of the Ice–Water Interface. I. Measurement of the Surface Potential. J*. Colloid Interface Sci*. **2000**, *232*, 81-85.

[22] Kallay, N.; Cop, A.; Chibowski, E.; Holysz, L. Reversible Charging of the Ice–Water Interface. II. Estimation of Equilibrium Parameters. *J. Colloid Interface Sci*. **2003**, *259*, 89-96.

[23] Workman, E. J.; Reynolds, S. E. Electrical Phenomena Occurring During the Freezing of Dilute Aqueous Solutions and their Possible Relationship to Thunderstorm Electricity. *Phys. Rev*. **1950**, *78*, 254-259.

[24] Wilson, P. W.; Haymet, D. J. Workman-Reynolds Freezing Potential Measurements between Ice and Dilute Salt Solutions for Single Ice Crystal Faces. *J. Phys. Chem. B*. **2008**, *112*, 11750-11755.

[25] Wilson, P. W.; Haymet, D. J. Effect of Ice Growth Rate on the Measured Workman-Reynolds Freezing Potential between Ice and Dilute NaCl Solutions. *J. Phys. Chem. B*. **2010**, *114*, 12585-12588.

[26] Lide, D. R. Handbook of Chemistry and Physics, 83[th] edition; CRC Press LLC: Boca

Raton, Florida, 1998

[27] Wilen, L. A.; Wettlaufer, J. S.; Elbaum, M.; Schick, M. Dispersion-Force Effects in Interfacial Premelting of Ice. *Phys. Rev. B*. **1995**, *52*, 12426.

[28] Barisik, M.; Atalay, S.; Beskok, A.; Qian, S. Size Dependent Surface Charge Properties of Silica Nanoparticles. *J. Phys. Chem. B.* **2014**, *118*, 1836-1842.

**SUPPORTING INFORMATION**

Expression of the forces used to compute the critical velocity:

**1. The disjoining force (between the particle and the ice front):**

$$F_d = -\frac{A_{p-i}R}{6(1 \pm \alpha)h_0^2}$$

where $A_{p-i}$ is the Hamaker constant between the ice water front and the silica particle within the liquid water medium; $R$ the nanoparticle radius and $h_0$ the minimum value of the distance between the nanoparticle and the solidifying front.

**2. The viscous force:**

$$F_\mu = -\frac{6\pi \beta v_p R^2}{(1 \pm \alpha)^2 h_0}$$

where $\mu$ is the viscosity of water; $v_p$ the velocity of the particle; $\beta = \frac{v_p}{v_f} = \frac{\rho_L - \rho_S}{\rho_L}$: a non-dimensional parameter derived from the boundary [16]; $v_f$ the velocity of the front; and $\rho_L$, $\rho_S$ are respectively the density of the liquid water and of the solid water.

**3. The DLVO force:**

$$F_{el} = -\frac{64\pi k_B T R \rho_0 \gamma^2}{\kappa} e^{-\kappa x} - \frac{A_{p-p}R}{6x^2}$$

where $k_B$ is the Boltzmann constant; $T$ the temperature; $\rho_0$ the charge density; $\gamma = \tanh\left(\frac{e\psi_0}{4k_B T}\right)$: the ratio between electric and thermal energy; $\kappa$ the inverse Debye length; $A_{p-p}$ the Hamaker constant between two particles within the liquid water medium; and $x$ the inter particles distance.

**4. The buoyancy force:**

$$F_g = -\frac{4}{3}\pi R^3 (\rho_P - \rho_L) g$$

where $\rho_P$ is the density of the particle. $F_g$ is negligible for nanoparticles.

**5. The electrostatic force between charged particle and charged ice front:**

The interaction energy between a charge sphere and a charge surface is [20]:

$$G = C_1 \left[ C_2 \left( h_0 - \frac{1}{2\kappa} \ln(e^{2h_0\kappa} - 1) \right) + \frac{\sigma_S \sigma_P}{\kappa} \ln \left( \frac{1 + e^{-h_0\kappa}}{1 - e^{-h_0\kappa}} \right) \right]$$

where $C_1 = \frac{2\pi R}{\kappa \epsilon_0 \varepsilon_W}$; $C_2 = \sigma_S^2 + \sigma_P^2$; and $h_0$ is the distance between the nanoparticle and the ice front.

Hence, the force $F_{el}^c$ is given by:

$$F_{el}^c = -\frac{\partial G}{\partial h_0}$$

$$F_{el}^c = \frac{C_1}{e^{2h_0\kappa} - 1} (C_2 + 2\sigma_S \sigma_P e^{h_0\kappa})$$

The critical velocity $v_c$ is reached when the velocities of the particle and the ice front are the same:

$$v_p = v_f \rightarrow \sum_i F_i = 0$$